\documentclass[aps,epjc,twocolumn,floatfix,superscriptaddress,nofootinbib,showpacs,showkeys,preprintnumbers]{revtex4}
\usepackage{graphicx}
\usepackage{dcolumn}  % Align table columns on decimal point
\usepackage{bm}       % bold math
\usepackage{palatino}
\preprint{preprint number}
\usepackage{epsfig}
\begin{document}
\title{ Possible scaling behaviour of the multiplicities ratio in 
leptoproduction of charged pions in
nuclear medium}
\author{N.~Akopov}
\author{L.~Grigoryan}
\author{Z.~Akopov \\  
Yerevan Physics Institute, Br.Alikhanian 2, 375036 Yerevan, Armenia }
%%%%%%%%%%%%%%%  begin abstract  %%%%%%%%%%%%%%%%%%%%%%%%%%%%%%%.ps&
\begin {abstract}
\hspace*{1em}
In this paper it is demonstrated that based on two-dimensional 
distributions in semi-inclusive deep
inelastic scattering (SIDIS) data, obtained recently by the HERMES 
experiment at DESY on different nuclei, which contains data for charged 
pions produced in $z$ slices as a function of $\nu$, and in a $\nu$ 
slices as a function of $z$, 
%%%% where $\nu$ and $z$ are the energy of the 
%%%%virtual photon and the fraction of that energy carried by the final 
%%%%hadron, respectively, 
it it is possible to parametrise a ratio of multiplicities on
nucleus and deuterium (per nucleon) $R_M^{h}$ in a form of a function 
of a single variable $\tau$, which has the physical meaning of the 
formation time of hadron. We call this effect $\tau$ scaling. $\tau$
is a function of two variables $\nu$ and $z$. It is also shown that 
$R_M^{h}$ can be presented in a form of a linear polynomial of $\tau$,
$a_{11}$ + $\tau a_{12}$, where parameters $a_{11}$ and $a_{12}$ do not 
depend on $\nu$ and $z$. 
\end {abstract}
\maketitle
%%%%%%%%%%%%%%%  end abstract  %%%%%%%%%%%%%%%%%%%%%%%%%%%%%%%
%%%%%%%%%%%%%%%  begin \section{Introduction}  %%%%%%%%%%%%%%%%%%%%%%%%%
\section{Introduction}
\normalsize
\hspace*{1em}
Hadronization is the process through which partons, created in an
elementary interaction, turn into the hadrons. Experimentally the 
hadronization process in free space (vacuum) has been studied 
extensively in $e^{+}e^{-}$ annihilation and in semi-inclusive
lepton-proton deep inelastic scattering (DIS). As a result, the spectra
of hadrons produced and their kinematical dependences are rather well
known. However, little is known about the space-time evolution of the 
process, because the products of this process can only be observed in
a detector that is separated from the reaction point by a macroscopic
distance. It is worth to mention that according to theoretical
estimates the hadronization process occurs over length scales that vary
from less than a femtometer to several tens of femtometers. The nuclear
medium can serve as a detector located directly at the place where 
microscopic interaction happens. Consequently, leptoproduction of 
hadrons on atomic nuclei provides a way to investigate the space-time
picture of the hadronization process. The semi-inclusive DIS of leptons
on nuclear targets is widely used for the study of this 
process~\cite{osborn,ashman,airap1,airap2}. It is most effective to
observe at moderate energies of the virtual photon, when the formation
time of the hadron is comparable with the nuclear radius. Such 
possibility provides HERMES experiment at DESY, which uses electron
(positron) beam with energy $27.5 GeV$ and fixed nuclear targets.\\
\hspace*{1em}
The most convenient observable measured experimentally for this process
is the nuclear attenuation ratio, which is a ratio of multiplicities on
nucleus and deuterium (per nucleon) for a given hadron. We shall denote
it as $R_M^{h}$. For a more profound understanding of the hadronization 
mechanism, it is important to find a variable which allows to present 
this observable in the most simple functional form.\\
\hspace*{1em}
Usually it is supposed, that $R_M^{h}$ is a function of two variables
$\nu$ and $z$, which are the energy of photon and the fraction of this
energy carried by the final hadron with energy $E_h$ ($z=E_h/\nu$),
respectively\footnote{In fact, $R_M^{h}$ also depends on the photon
virtuality $Q^2$ and on the square of the  hadron transverse momentum in
respect to the virtual photon direction, $p_t^{2}$. However, from the
experimental data, it is known that $R_M^{h}$ is a much sensitive function
of $\nu$ and $z$ in comparison with $Q^2$ and $p_t^{2}$.}.\\
\hspace*{1em}
In our preceding work~\cite{simpl} we performed a fit for the evidence 
that the formation time $\tau$ is the best variable for $R_M^{h}$, i.e. 
that it 
can be parametrized as a function of a single variable $\tau$. Three 
widely known representations for $\tau$ were used for the fit. The
experimental data for pions on nitrogen and for identified hadrons on
krypton nuclei obtained by the HERMES experiment~\cite{airap1,airap2} 
were used for this fit.
We have demonstrated that the nuclear attenuation ratio can be 
presented, with good
precision, as a function of a single variable $\tau$ instead of a 
function of two variables $\nu$ and $z$. Moreover,  
$R_M^{h}$ is a linear function of $\tau$. We named $\tau$ a scaling 
variable, because
it contains all $\nu$ and $z$ dependencies of $R_M^{h}$. For the fit we 
have obtained $R_M^{h}$ as a function of $\tau$ from experimental
data~\cite{airap1,airap2}, where it was measured as a function of $\nu$ with the 
integration over $z$, and as a function of $z$ with the integration over $\nu$. This means,
that the data were taken at an unequal binning over $\nu$ and $z$. In 
case of
$\nu$-dependence the detailed bins over the variable $\nu$ were taken, and for each
value of $\nu$ the value of $z$ averaged over whole range of measured
$z$ ($<z>$) was taken. In case of $z$-dependence we also performed the above mentioned procedure.\\
%to get a detailed bins over variable
%$z$ and for each value of $z$ was taken value of $\nu$ averaged over all
%region of measured $\nu$, $<\nu>$.\\
\hspace*{1em} 
Recent work of published by the HERMES experiment~\cite{airap3} allows 
to escape this 
difficulty,  because the data published contains, among 
others, the so called  two-dimensional data, i.e. multiplicity ratio 
$R_M^{h}$ for charged pions 
produced in a $z$ slices as a function of $\nu$, and in a $\nu$ slices as a 
function of $z$. The data were obtained for four nuclear targets and 
used 
for a new fit of $R_M^{h}$ as a function of $\tau$.\\
\hspace*{1em}
The main aim of this work is to show, that $R_M^{h}$ is a function of 
single variable 
$\tau$, rather than a function of two variables $\nu$ and $z$, using 
the new set of $R_M^{h}$ given by the two-dimensional analysis, where 
the data is split into more regular $\nu$ and $z$ bins than in case of 
traditional presentation in form of $\nu$ - and
$z$ - dependencies, 
This will allow to verify the results of our preceding work~\cite{simpl} 
in more favourable conditions, i.e. to confirm that in electroproduction 
of hadrons in nuclear medium we indeed observe scaling, where $\tau$
takes over the role of the scaling variable.\\
\hspace*{1em}
This paper is organized as follows. Nuclear attenuation in an absorption   
model is presented in the next section. In Section 3 we discuss the choice
of an appropriate form for the variable $\tau$. Section 4 presents results
of the fit. Conclusions are given in Section 5.
%%%%%%%%%%%%%%%  end \section{Introduction}  %%%%%%%%%%%%%%%%%%%%%%%%%
%%%%%%%%%%%%%%%  begin \section{NA in absorption model}  %%%%%%%%%%%%%%%
\section{Nuclear attenuation in absorption model}
\normalsize
\hspace*{1em}
The semi-inclusive DIS of lepton on nucleus of atomic mass number A is
\begin{eqnarray}
          l_i + A \rightarrow l_f + h + X,
\end{eqnarray}
where $l_i (l_f)$ and $h$ denote the initial (final) leptons and the hadron
observed in the final state. This process is usually investigated in
terms of $R_M^{h}$ :
\begin{eqnarray}
       {R_M^{h}(\nu,z) = 2d\sigma_A(\nu,z)/Ad\sigma_D(\nu,z)}.
\end{eqnarray}
Experimental data are usually presented at precise values of one
variable and average values of another\footnote{In case where the
$\nu$ - dependence is studied, we have $R_M^{h}(\nu,\langle z \rangle)$,
where $\langle z \rangle$ are the average values of $z$ for each $\nu$
bin. For $z$ - dependence $R_M^{h}(\langle \nu \rangle ,z)$,where
$\langle \nu \rangle$ are the average values of $\nu$ for each $z$ bin.}.
\hspace*{1em}
In this work we adopt a model according to which the origin of the
nuclear attenuation is the absorption of the prehadron (string, dipole)
and final hadron in the nuclear medium. In that case $R_M^{h}$ has the
following form:
\begin{eqnarray}
{R_M^{h}=\int{d^2b} \int_{-\infty}^{\infty}
{\rho(b,x)[W(b,x)]^{(A-1)}dx}},
\end{eqnarray}
where $b$ is the impact parameter and $x$ the longitudinal coordinate of
the DIS point. $\rho$ is the nuclear density function with a normalization
condition $\int{\rho(r)d^{3}r}=1$. $W(b,x)$ is the probability that
neither the prehadron nor the final hadron $h$ are absorbed by a nucleon
located anywhere in the nucleus. For $W(b,x)$ we use the one-scale model
proposed in Ref.~\cite{bialas}:
\begin{eqnarray}
\nonumber
{W(b,x)=1-\sigma_q \int_{x}^{\infty}{P_q(x'-x)\rho(b,x')dx'}} \\
{-\sigma_h \int_{x}^{\infty}{P_h(x'-x)\rho(b,x')dx'}},
\end{eqnarray}
where $\sigma_q$ and $\sigma_h$ are the inelastic cross sections for
prehadron-nucleon and hadron-nucleon interactions, respectively. Generally 
speaking $\sigma_q$ is a function of the distance $x'-x$, $\nu$ and $z$
\footnote{$\sigma_q$ is a function of formation time $\tau$ rather than 
a function of variables $\nu$ and $z$ separately.}.
However a comparison of simple theoretical models containing $\sigma_q$
as a parameter~\cite{ashman,akopov}, with the experimental data obtained in
different kinematical regions, (in particular in different domains of $\nu$),
shows that the approximation $\sigma_q = const$ leads to a very acceptable
agreement with the data. Further, taking into account the qualitative
character of this work, we shall use this approximation. In the region of
moderate energies, $\sigma_h$ are approximately constant for all $h$.
$P_q(x'-x)$ is the probability that at distance $x'-x$ from the DIS point,
the particle is a prehadron and $P_h(x'-x)$ is the probability that the
particle is a hadron. The abovementioned probabilities are related via a
condition
\begin{eqnarray}
{P_h(x'-x)=1-P_q(x'-x)}.
\end{eqnarray}
In analogy with the survival probability for a particle having lifetime
$\tau$ in a system where it travels a distance $x'-x$ before decaying,
$P_q(x'-x)$ can be expressed in the form
\begin{eqnarray}
{P_q(x'-x)=\exp[-(x'-x)/\tau]}, 
\end{eqnarray}  
where $\tau$ is the formation time. Substituting expressions for $P_q(x'-x)$
and $P_h(x'-x)$ in eq.(4) one obtains
\begin{eqnarray}
\nonumber
{W(b,x)\approx{1-\sigma_h \int_{x}^{\infty}{\rho(b,x')dx'}
+{\tau(\sigma_h-\sigma_q)\rho(b,x)}}}\\
{\approx{w_1(b,x)+\tau(\nu,z)w_2(b,x)}.\hspace{2.4cm}}
\end{eqnarray}
In the framework of our suppositions about $\sigma_q$ and $\sigma_h$,
$W$ depends on $\nu$ and $z$ only by means of $\tau(\nu,z)$.\\
\hspace*{1em}
In more detail, the formation time in the string models can be divided 
in
two parts (see, for instance, the two scale model presented in 
Refs.~\cite{ashman,akopov}). The first part is the constituent formation
time $\tau_c$, which defines the time elapsed from the moment of the DIS
untill the production of the first constituent of the final hadron. The
second time interval begins with the production of the first constituent
until the second one, which coincides with the yo-yo\footnote{The yo-yo
formation means that a colorless system with valence contents and
quantum numbers of the final hadron is formed, but without its "sea"
partons.} or final hadron production. Comparison with the experimental
data shows that in the second interval, the prehadron-nucleon cross section
has values close to the hadron-nucleon cross section $\sigma_h$. If the
difference between these cross-sections is neglected, the model is reduced
to one scale model with $\tau = \tau_c$. Substituting $W(b,x)$ in $R_M^{h}$
we obtain
\begin{eqnarray}
\nonumber 
{R_M^{h} \approx{\int{d^2b} \int_{-\infty}^{\infty}{\rho(b,x)
(w_1+\tau w_2)^{(A-1)}dx}}}\\
\approx{a_{i1} + \tau a_{i2} + \tau^2 a_{i3} + {}\cdots},\hspace{1.85cm}
\end{eqnarray}
where $i$ is the maximal power of $\tau$ with which we are limited.
Although $R_M^{h}$ is a polynomial of $\tau$ with maximal power $A-1$,
it is expected that $a_{i1} > a_{i2} > a_{i3} > {}\cdots$. The
coefficients $a_{ij}$ depend on $A, \sigma_q, \sigma_h$ and nuclear
density. For fit we use three expressions for $R_M^{h}$ as first, second
and third order polynomials of $\tau$ :
\begin{eqnarray}
{R_M^{h}[P_1]={a_{11} + \tau a_{12} },\hspace{2.35cm}}
\end{eqnarray}
\begin{eqnarray}
{R_M^{h}[P_2]={a_{21} + \tau a_{22} + \tau^2 a_{23} },\hspace{1.1cm}}
\end{eqnarray}  
\begin{eqnarray}
{R_M^{h}[P_3]={a_{31} + \tau a_{32} + \tau^2 a_{33} + \tau^3 a_{34}}}.
\end{eqnarray}
In order to get the information on the influence of highest order 
polynomial forms for $R_M^{h}$, $R_M^{h}[P_4]$ expression also was checked 
(see section 4).
%%%%%%%%%%%%%%%  begin \section{Formation time}  %%%%%%%%%%%%%%%
\section{Formation time}
\normalsize
\hspace*{1em}
Equation (8) shows that within our approximation, $R_M^{h}$ depends on
$\nu$ and $z$ only by means of $\tau(\nu,z)$. This is the reason why we 
call $\tau$ a scaling variable. In this section we shall discuss the 
physical meaning and possible expressions of the formation time $\tau$. 
There are different definitions for the formation time. We define it as 
a time scale which is necessary for the prehadron-nucleon 
cross section to reach the value of the hadron-nucleon cross-section. In 
the literature there are three qualitatively different definitions for 
$\tau$. In the first extreme
case it is assumed that $\tau = 0$ (Glauber approach). In the second
extreme case $\tau \gg r_A$, where $r_A$ is the nuclear radius (energy
loss model~\cite{energyloss}). And at last, and in our opinion more 
realistic definition of the formation time, as a function of $\nu$ and 
$z$ which can change from zero up to values larger than $r_A$. 
Experimental data seem to confirm that for moderate values of $\nu$ (on 
the order of $10 GeV$) the
formation time is comparable with the nuclear size, i.e. the hadronization
mostly takes place within the nucleus. This follows from the comparison
of the experimental data for $R_M^{h}$ obtained in the region of 
moderate~\cite{airap2} and high~\cite{ashman} energies. At moderate
energies $R_M^{h}$ significantly differs from unity and is a sensitive
function of $\nu$ and $z$, at high energies $R_M^{h} \approx 1$ and weakly
depends on $\nu$ and $z$. For the formation time we shall use expressions
which do not contradict the third definition mentioned above. The following 
expressions are used:\\
1. Formation time for the leading hadron~\cite{kopel}, which follows from
the energy-momentum conservation law
\begin{eqnarray}
{\tau_{lead.}=(1-z)\nu/\kappa} ,
\end{eqnarray}
where $\kappa$ is the string tension (string constant) with numerical value
$\kappa = 1 GeV/fm$.\\
2. Formation time for the fast hadron, which is composed of characteristic 
formation time of the hadron $h$ in its rest frame $\tau_0$ and Lorentz
factor (see, for instance, Ref.~\cite{bialas})
\begin{eqnarray}
{\tau_{Lor.}=\tau_0\frac{E_h}{m_h}=\tau_0\frac{z\nu}{m_h}} , 
\end{eqnarray}
where $E_h$ and $m_h$ are the energy and mass of the hadron $h$, 
respectively. In Ref.~\cite{simpl} we have discussed formation time 
$\tau_{Lor.}$
in detail. In particular we considered the possibility of that $\tau_0$ 
being proportional to $m_h$. In the present paper we have to deal with 
hadrons of one type only - charged pions - and looking at interested 
$\nu$ and $z$ dependencies of $\tau$, because all factors, which do not 
depend from these variables, can only renormalize coefficients $a_{ij}$ 
(see equations (9)-(11)).\\
3. The formation time following from the Lund string model in
Ref.~\cite{Bi_Gyul} is\footnote{Note that this approximation is used only
for the sake of convenience. For numerical calculations we use the precise
expression for $\tau_{Lund}$ following from equation 
$\tau_{Lund}=\tau_y-z\nu/\kappa$ with $\tau_y$ taken from eq.(4.21) of 
Ref.~\cite{Bi_Gyul}.}
\begin{eqnarray}
{\tau_{Lund}=\Bigg[{\frac{\ln(1/z^2)-1+z^2}{1-z^2}}  
\Bigg]{\frac{z\nu}{\kappa}}} .
\end{eqnarray}
\hspace*{1em}
One should note that all three types of formation time have similar behavior
with $\nu$, but different behavior with $z$. At the values of $z$ typical
for the HERMES kinematics ($z \geq 0.2$) the behavior of $\tau$ defined as in 
eqs.(12) and (14) with $z$ is similar, i.e. they are decreasing with the
increase of $z$, while $\tau$ defined as in eq.(13) is increasing with the
increase of $z$.
%%%%%%%%%%%%%%%  end \section{Formation time}  %%%%%%%%%%%%%%%  
%%%%%%%%%%%%%%%  begin \section{Results}  %%%%%%%%%%%%%%%
\section{Results}
\normalsize
\hspace*{1em}
The two-dimensional data from~\cite{airap3}, i.e. the multiplicity ratio
$R_M^{h}$ for charged pions produced on helium, neon, krypton and xenon
nuclei, in a $z$ slices as a function of $\nu$, and in a $\nu$ slices as a
function of $z$ were used to perform the fit.
%- begin FIGURE -----------------------------------------------------
\begin{figure}[!ht]
\begin{center}
\epsfxsize=8.cm 
\epsfysize=10.cm 
\epsfbox{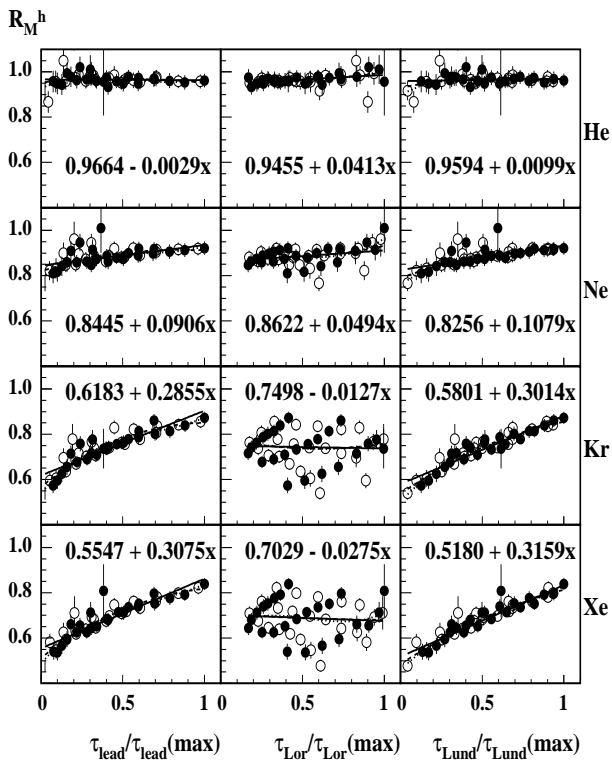}
\end{center}
\caption{\label{xx1}
{\it Multiplicity ratios for charged pions on different nuclei as a
function of
$\tau_{lead.}$ (left panels),
$\tau_{Lor.}$ (central panels),
$\tau_{Lund}$ (right panels).
Normalized values $x = \tau/\tau(max)$ for all $\tau$ are  used.
On panels from upper to lower results for helium, neon, krypton
and xenon nuclei are presented, respectively. Solid, dashed and
dotted curves are results of linear, quadratic and cubic polynomial
fits. The numerical results for the linear fit presented on the panels.
Experimental points obtained from~\cite{airap3}. Details see in text.
}}
\end{figure}
%- end FIGURE -------------------------------------------------------
The independent fit including 47 experimental points was performed for
each nucleus. As it is clear from eq.(2), experimental points corresponding
to $\nu$ - dependence $R_M^{h}(\nu,\langle z \rangle)$, and $z$ - dependence
$R_M^{h}(\langle \nu \rangle ,z)$ enter in fit on equal basis, as a
values of function $R_M^{h}(\nu ,z)$ at values of variables $(\nu ,z)$
equal $(\nu,\langle z \rangle)$ and $(\langle \nu \rangle ,z)$, respectively. 
For the fit $R_M^{h}$ has been taken in polynomial forms $R_M^{h}[P_{1,2,3}]$
[see eqs.(9)-(11)], and formation times (lengths) as in eqs.(12)-(14). The
results for the reduced $\chi^2$ denoted as $\chi^2/{d.o.f.}$ presented in
Table~\ref{tab:Table1}. One can see that for each choice of formation time
and for each nucleus, the values of $\chi^2/{d.o.f.}$ are close for the
polynomial approximations $R_M^{h}[P_1]$, $R_M^{h}[P_2]$, and $R_M^{h}[P_3]$,
which means that the inclusion in consideration of the higher order
polynomials of $\tau$ does not essentially improve the description of the
data.\\
\hspace*{1em}
In order to test this, we have also calculated the $R_M^{h}[P_4]$ polinomial
form and obtained the values of $\chi^2/{d.o.f.}$ close to the ones in case
of $R_M^{h}[P_3]$.
%--- begin TABLE --------------------------
%\onecolumn
%\vspace {-1cm}
%\Large
\begin{table*}[!ht]
\normalsize
\caption{ The $\chi^2/{d.o.f.}$ values obtained from polynomial fit. 
$P_{1,2,3}$ denote the expressions $R_M^{h}[P_{1,2,3}]$ used as fitting
functions. The necessary details concerning the data sets used for the
fit are given in the text.}
\label{tab:Table1}
\begin{center}
\begin{tabular}{  c  c  c  c  c  c  c  c  c  c  c  c  }
\hline
\hline\\[0.0cm]
\multicolumn{3}{c}{}&\multicolumn{3}{c}{$\tau_{lead.}$}&
\multicolumn{3}{c}{$\tau_{Lor.}$}&\multicolumn{3}{c}{$\tau_{Lund}$}\\[0.1cm]
%\hline
\cline{5-5}
\cline{8-8}
\cline{11-11}\\[0.0cm] 
A&$N_{exp}$&$Had.$
&\hspace{0.35cm}$P_1$\hspace{0.35cm}
&\hspace{0.35cm}$P_2$\hspace{0.35cm}
&\hspace{0.35cm}$P_3$\hspace{0.35cm}
&\hspace{0.35cm}$P_1$\hspace{0.35cm}
&\hspace{0.35cm}$P_2$\hspace{0.35cm}
&\hspace{0.35cm}$P_3$\hspace{0.35cm}
&\hspace{0.35cm}$P_1$\hspace{0.35cm}
&\hspace{0.35cm}$P_2$\hspace{0.35cm}
&\hspace{0.35cm}$P_3$\hspace{0.35cm}\\
\hline
{\hspace{0.3cm}$^{4}He$}&{47}&{$<\pi>$}
&\hspace{0.35cm}{0.43}\hspace{0.35cm}
&\hspace{0.35cm}{0.42}\hspace{0.35cm}
&\hspace{0.35cm}{0.39}\hspace{0.35cm}
&\hspace{0.35cm}{0.37}\hspace{0.35cm}
&\hspace{0.35cm}{0.37}\hspace{0.35cm}
&\hspace{0.35cm}{0.38}\hspace{0.35cm}
&\hspace{0.35cm}{0.42}\hspace{0.35cm}
&\hspace{0.35cm}{0.39}\hspace{0.35cm}
&\hspace{0.35cm}{0.36}\hspace{0.35cm}\\
%\hline
{\hspace{0.2cm}$^{20}Ne$}&{47}&{$<\pi>$}
&\hspace{0.35cm}{0.62}\hspace{0.35cm}
&\hspace{0.35cm}{0.55}\hspace{0.35cm}
&\hspace{0.35cm}{0.50}\hspace{0.35cm}
&\hspace{0.35cm}{0.98}\hspace{0.35cm}
&\hspace{0.35cm}{0.98}\hspace{0.35cm}
&\hspace{0.35cm}{0.96}\hspace{0.35cm}
&\hspace{0.35cm}{0.45}\hspace{0.35cm}
&\hspace{0.35cm}{0.38}\hspace{0.35cm}
&\hspace{0.35cm}{0.35}\hspace{0.35cm}\\
%\hline
{\hspace{0.2cm}$^{84}Kr$}&{47}&{$<\pi>$}
&\hspace{0.35cm}{1.27}\hspace{0.35cm}
&\hspace{0.35cm}{0.83}\hspace{0.35cm}
&\hspace{0.35cm}{0.76}\hspace{0.35cm}
&\hspace{0.35cm}{7.18}\hspace{0.35cm}
&\hspace{0.35cm}{7.31}\hspace{0.35cm}
&\hspace{0.35cm}{7.47}\hspace{0.35cm}
&\hspace{0.35cm}{0.73}\hspace{0.35cm}
&\hspace{0.35cm}{0.57}\hspace{0.35cm}
&\hspace{0.35cm}{0.50}\hspace{0.35cm}\\
%\hline
{\hspace{0.1cm}$^{131}Xe$}&{47}&{$<\pi>$}
&\hspace{0.35cm}{0.83}\hspace{0.35cm}
&\hspace{0.35cm}{0.55}\hspace{0.35cm}
&\hspace{0.35cm}{0.50}\hspace{0.35cm}
&\hspace{0.35cm}{7.68}\hspace{0.35cm}
&\hspace{0.35cm}{7.85}\hspace{0.35cm}
&\hspace{0.35cm}{7.98}\hspace{0.35cm}
&\hspace{0.35cm}{0.54}\hspace{0.35cm}
&\hspace{0.35cm}{0.45}\hspace{0.35cm}
&\hspace{0.35cm}{0.42}\hspace{0.35cm}\\
\hline
\hline
\end{tabular}
\end{center}
\end{table*}
%\twocolumn
\normalsize
%--- end TABLE ----------------------------
From Table~\ref{tab:Table1} one can see that the fit gives
unexpectedly good values for $\chi^2/{d.o.f.}$ in case of $\tau_{lead.}$ and
$\tau_{Lund}$, for $\tau_{Lor.}$ the agreement is much worse. As it is known
from experiment~\cite{airap1, airap2, airap3}, 
$R_M^{h}(\nu,\langle z \rangle)$ increases with increasing of $\nu$, and
$R_M^{h}(\langle \nu \rangle ,z)$ decreases with increasing of $z$ for all
nuclei. Our assumption is that these functions indeed present different
representations of the same function, which depends {\it on  
variable
$\tau$ only}. In turn $\tau$ is a function of $\nu$ and $z$. Now let us
discuss the figure. Using eqs.(12)-(14), we present
$R_M^{h}(\nu,\langle z \rangle)$ and $R_M^{h}(\langle \nu \rangle ,z)$ as
functions of $\tau$. Experimental points and results of the fit are presented
in Fig.~\ref{xx1}. Solid points correspond to the $R_M^{h}(\nu,\langle z \rangle)$
obtained from the experimental data for $\nu$-dependence, open points to the
$R_M^{h}(\langle \nu \rangle ,z)$ from $z$-dependence. From the figure one can
easily note that experimental points for $R_M^{h}(\nu,\langle z \rangle)$ and 
$R_M^{h}(\langle \nu \rangle ,z)$ as functions of $\tau$ have the same behavior
and approximately coincide when $\tau_{lead.}$ and $\tau_{Lund}$ serve as the
variables. The reason for this is that these variables are approximately
proportional to $\nu$ and $1 - z$. In contrary, variable $\tau_{Lor.}$ is
proportional to $\nu$ and $z$, and as a consequence
$R_M^{h}(\nu,\langle z \rangle)$ and $R_M^{h}(\langle \nu \rangle ,z)$ have
opposite behavior as functions of $\tau_{Lor.}$. This means, that without any
calculations one can state that $\tau_{lead.}$ and $\tau_{Lund}$ can serve as
a scaling variables, but $\tau_{Lor.}$ cannot. For the sake of convenience we
have renormalized $\tau$ to $x = \tau/\tau(max)$, where $\tau(max)$ are the
maximum values of $\tau$ for each set of data and each choice of the $\tau$
expression.\\
\hspace*{1em}
The range of variation of $\tau$ and the numerical values for $\tau(max)$ in
all scenarios for instance on krypton are:
$2.46 \leq \tau_{Lor.} \leq 14.6 fm$,
$0.37 \leq \tau_{lead.} \leq 15.5 fm$,
$0.36 \leq \tau_{Lund} \leq 8.28 fm$.
Presentation of $R_M^{h}$ as a function of $x$ allows us to place all data in
an interval $(0, 1)$. This choice does not influence the results of the fit and
the values of $R_M^{h}$. On the figure the linear polynomial is presented
$a_{11} + x a_{12}^{'}$ with values $a_{11}$ and
$a_{12}^{'}$ = $a_{12}\tau(max)$ corresponding to the best fit. Solid, dashed
and dotted curves represent the $R_M^{h}[P_1]$, $R_M^{h}[P_2]$, and
$R_M^{h}[P_3]$ polynomial fit, respectively. One can easily see that the
difference between the curves corresponding to $R_M^{h}[P_1]$, $R_M^{h}[P_2]$,
and $R_M^{h}[P_3]$ is small. The vertical positions of the experimental points
are the same in all scenarios. The experimental points can be closer together
(or not) depending on the type of the formation time definition. When looking
at the $\tau$ dependencies, the points corresponding to $\nu$-dependence
preserve their order in all scenarios. In case of $z$-dependence, the order is
the same in the scenario with $\tau_{Lor.}$, but changes to opposite in other
scenarios.\\
\hspace*{1em}
As a last remark, one should note that results of this analysis do not depend
from the values of the parameters, in particular from the values of $\kappa$
and $\tau_{0}$.
%%%%%%%%%%%%%%%  end \section{Results}  %%%%%%%%%%%%%%%
%%%%%%%%%%%%%%%  begin \section{Conclusions}  %%%%%%%%%%%%%%%
\section{Conclusions.}
\normalsize
\hspace*{1em}
The two-dimensional nuclear attenuation data for charged pions on helium, neon,
krypton and xenon nuclei obtained recently by the HERMES experiment~\cite{airap3} 
were used to perform the fit. So far it has been supposed that 
experimentally measured function $R_M^{h}(\nu,\langle z \rangle)$ 
depends from variable $\nu$ only, and $R_M^{h}(\langle \nu \rangle ,z)$ 
from $z$ only. In our preceding work~\cite{simpl}, we have assumed that 
these functions indeed present different
representations of the same function, which depends on only one 
variable $\tau$.
In turn $\tau$ is a function of $\nu$ and $z$. In this work based on 
two-dimensional distributions (more suitable for this investigation), 
the results of our preceding work~\cite{simpl} are confirmed. We 
demonstrate (see Table 1 and figure) that
$R_M^{h}(\nu,\langle z \rangle)$ and $R_M^{h}(\langle \nu \rangle ,z)$ as a
functions of $\tau$ have the same behavior and approximately coincide, when $\tau$ is used in form of
$\tau_{lead.}$ and $\tau_{Lund}$. In contrary, they have
opposite behavior as a functions of $\tau_{Lor.}$. This indicates that
$\tau_{lead.}$ and $\tau_{Lund}$ can serve as scaling variables, but
$\tau_{Lor.}$ cannot. We also show that $R_M^{h}(\tau)$, with a good precision, can be parametrised in a form of a linear polynomial 
$a_{11}$ + $\tau a_{12}$, where the fitting parameters $a_{11}$ and 
$a_{12}$ do not depend on $\nu$ and $z$.
We conclude that experimentally measured function $R_M^{h}$ is a 
function of single variable $\tau$, and that $\tau$ scaling 
follows naturally from the absorption model, but we do not confirm that 
scaling is a property of an absorption model only. There could exist 
other mechanisms which can also lead to the scaling behavior - this is 
an open question, which requires further investigation.
%%%%%%%%%%%%%%%  end \section{Conclusions}  %%%%%%%%%%%%%%%
%--- begin BIBLIOGRAPHY -------------------

%--- end BIBLIOGRAPHY ---------------------

\begin{thebibliography}{99}
\bibitem{osborn}  L.S.~Osborn et al., Phys. Lett. {\bf 40B} (1978) 1624
\bibitem{ashman}  J.Ashman et al., Z.Phys. {\bf C52} (1991) 1
\bibitem{airap1}  A.Airapetian et al., Eur.Phys. J. {\bf C20} (2001) 479
\bibitem{airap2}  A.Airapetian et al., Phys.Lett. {\bf B577} (2003) 37
\bibitem{simpl}  N.Akopov, L.Grigoryan, Z.Akopov, Phys.Rev.{\bf C76} (2007) 
065203; hep-ph/0703124 (2007)
\bibitem{airap3}  A.Airapetian et al., Nucl.Phys. {\bf B780} (2007) 1
\bibitem{bialas}  A.Bialas, Acta Phys.Polon. {\bf B11} (1980) 475
\bibitem{akopov}  N.Akopov, L.Grigoryan, Z.Akopov, Eur.Phys.J. {\bf C44}
(2005) 219; hep-ph/0409359 (2004)
\bibitem{energyloss} X.-N.~Wang, X.~Guo, Nucl. Phys., {\bf A696} (2001) 788;
E.~Wang, X.-N.~Wang, Phys. Rev. Lett. {\bf 89} (2002) 162301
\bibitem{kopel}  B.Kopeliovich, Phys.Lett. {\bf B243} (1990) 141
\bibitem{Bi_Gyul}  A.Bialas, M.Gyulassy, Nucl.Phys. {\bf B291} (1987) 793
\end{thebibliography}
\end{document}